\documentclass[%
 reprint,
 preprintnumbers,
 amsmath,amssymb,
 aps,
 prl,
]{revtex4-1}

\usepackage{empheq,epsfig}
\usepackage{graphicx}
\usepackage{dcolumn}
\usepackage{bm}

\usepackage{color}

\newcommand{\amp}{M}                     
\newcommand{\ampRid}{\mathfrak{M}}             
\newcommand{\bk}[1]{\langle #1 \rangle}  
\newcommand{\bs}{\boldsymbol}

\newcommand{\bt}{{\boldsymbol{b}}}

\newcommand{\dif}{\mathrm{d}}             

\newcommand{\el}{\mathrm{el}}           
\newcommand{\esp}[1]{\mathrm{e}^{#1}}    
\newcommand{\GW}{\mathrm{GW}}             
\newcommand{\hti}{\tilde{h}}

\newcommand{\ord}[1]{\mathcal{O}\left(#1\right)}

\newcommand{\pol}{\epsilon}               
\newcommand{\Qt}{{\boldsymbol{Q}}}
\newcommand{\qt}{{\boldsymbol{q}}}
\newcommand{\regge}{\mathrm{Regge}}
\newcommand{\soft}{\mathrm{soft}}
\newcommand{\tfa}{{\cal M}}               
\newcommand{\Tht}{{\boldsymbol{\Theta}}}
\newcommand{\tht}{{\boldsymbol{\theta}}}

\newcommand{\ui}{\mathrm{i}}             
\newcommand{\vp}{\vec{p}}                
\newcommand{\vq}{\vec{q}}                
\newcommand{\xt}{{\boldsymbol{x}}}
\newcommand{\zt}{{\boldsymbol{z}}}

\begin{document}

\title{Emerging 
Hawking-like
 Radiation from Gravitational Bremsstrahlung Beyond the Planck Scale}

\author{Marcello Ciafaloni}
\email{ciafaloni@fi.infn.it}
\author{Dimitri Colferai}
\email{colferai@fi.infn.it}
\affiliation{ Dipartimento di Fisica, Universit\`a di Firenze and INFN, Sezione di Firenze,
  Via Sansone 1, 50019 Sesto Fiorentino, Italy}
\author{Gabriele~Veneziano}
\email{gabriele.veneziano@cern.ch}
\affiliation{Coll\`ege de France, 11 place M. Berthelot, 75005 Paris, France}
\affiliation{Theory Division, CERN, CH-1211 Geneva 23, Switzerland}


\begin{abstract}
  We argue that, as a consequence of the graviton's spin-2, its bremsstrahlung
  in trans-planckian-energy ($E\gg M_P$) gravitational scattering at small
  deflection angle can be nicely expressed in terms of
  helicity-transformation phases and their transfer within the scattering
  process.  The resulting spectrum exhibits deeply sub-planckian characteristic
  energies of order $M_P^2/E \ll M_P$ (reminiscent of Hawking radiation), a
  suppressed fragmentation region, and a reduced rapidity plateau, in broad
  agreement with recent classical estimates.
\end{abstract}

\preprint{CERN-PH-TH-2015-131}

\maketitle

It is well known that much --- if not all --- of the geometric beauty of
Classical General Relativity can follow from assuming the existence, at the
quantum level, of a massless spin-2 particle, the graviton. This is how we
understand, for instance, that quantum string theory in flat space-time becomes
a theory of quantum (and, in some approximation, of classical) gravity, though
not necessarily Einstein's.

The emergence of a Schwarzschild metric through a resummation of
graviton-exchange diagrams was pointed out long ago by Duff \cite{Duff:1973zz}.
Much later, a similar approach was taken up in the context of string
theory~\cite{ACV87,ACV88,ACV89,GrMe87,GrMe88} where scattering at transplanckian
energy ($E\gg M_P \equiv\sqrt{\hbar/G}$, in $c=1$ units) was taken as the
thought experiment of choice for understanding quantum string gravity as well as
its quantum field theory and classical limits. It was possible to show
\cite{ACV87} how an effective Aichelburg-Sexl (AS) metric~\cite{AiSe71} emerges,
manifesting itself at large-impact-parameter ($b \gg R\equiv 4GE$) via the
gravitational deflection and tidal excitation \cite{ACV87,Giddings:2006vu} of
the incoming strings.

As one proceeds to smaller impact parameters corrections of relative order
$R^2/b^2$ appear~\cite{ACV90,ACV07}. These modify, of course, deflection angles
and time delays~\cite{CC14,Camanho:2014apa,D'Appollonio:2015gpa}, but also
introduce as a new phenomenon graviton bremsstrahlung.  At lowest-order all this
can be studied in terms of the so-called H-diagram~\cite{ACV90,ACV07}, but its
extension to higher orders turned out to be non trivial. In particular, the most
na\"ive resummation appears to endanger energy-conservation~\cite{Rychkov_pc}.

The purpose of this note is to go beyond the analysis of ~\cite{ACV90,ACV07} and
to show that the graviton's spin 2, besides making it possible for an effective
metric to emerge, also determines the detailed form of graviton bremsstrahlung
in a whole frequency and angular range, covering in particular the forward
fragmentation regions responsible for the excessive energy emission. As we will
show, taking properly into account coherence effects not only solves the
energy-conservation issue, but also leads to a gravitons spectrum with
characteristic energies of order $\hbar R^{-1}$, the typical energy of Hawking's
radiation out of a black hole of mass $E$.  The main features of such a picture
are consistent with their classical counterparts recently discussed
in~\cite{GrVe14, Spirin:2015wwa}.  In this short note we will sketch the
derivation and present the main physical results leaving most of the technical
details to a forthcoming paper~\cite{CCCV15}.

In order to set the framework consider first the elastic gravitational
scattering $p_1+p_2\to p'_1+p'_2$ of two fast particles, at center-of-mass
energy $2E=\sqrt{s}\gg M_P$, and momentum transfer $Q^\mu$ with transverse
component $\Qt\equiv E\Tht_s$, where the 2-vector
$\Tht_s=|\Tht_s|(\cos\phi_s,\sin\phi_s)$ describes both azimuth $\phi_s$ and
polar angle $|\Tht_s| \ll 1$ of the final particles with respect to the
longitudinal $z$-axis. This elastic scattering is described by the semiclassical
$S$-matrix $~\exp(2\ui\delta)$ whose leading term is given by the eikonal function
$\delta_0=(Gs/\hbar)\log(L/|\bt|)$ ($L$ being an irrelevant infrared cutoff).
Its exponentiation describes the amplitude at impact
parameter $\bt$, conjugated to $\Qt$, as a sum over a large number
$\bk{n}\sim Gs/\hbar$ of single-hit processes provided by single-graviton
exchanges of momenta $\qt_j=E\tht_j$ ($j=1,\cdots,n$).  The single-hit
scattering angle is very small, of order $\theta_m\equiv\hbar/Eb$, while the
overall scattering angle --- though small for $b \gg R$ --- is much larger, of
order $2(Gs/\hbar)\theta_m\simeq 2R/b$, the Einstein deflection angle.

In order to compute the emission amplitude of a graviton of momentum
$q^\mu = \hbar\omega(1,\tht,\sqrt{1-\tht^2})$, we start by considering the
single-exchange scattering amplitude of momentum transfer $\qt_s=E\tht_s$
($\theta_s\sim\hbar/Eb\equiv\theta_m$) and discuss various angular regimes,
under the assumption that the emitted graviton energy $\hbar\omega\ll E$. That
restriction still allows for a huge graviton phase space, in which classical
frequencies of order $R^{-1}$ --- and even much larger ones --- are available,
due to the large gravitational charge $\alpha_G \equiv Gs/\hbar \gg1$. We will
distinguish three regimes:

a) $|\tht_s|>|\tht|$. In this regime, characterized by small emission angles and
subenergies, the emission amplitude is well described by external-line
insertions corresponding to the Weinberg current, but the collinear limit has no
  singularities because of helicity-conservation zeroes.

b) $|\tht|>|\tht_s|>\frac{\hbar\omega}{E}|\tht|$. In this regime the
sub-energies reach the threshold of high-energy Regge behavior, still
remaining in the validity region of external-line insertions, due to the
condition $|\qt_s|>|\qt|$ which suppresses  the emission from the
exchanged-graviton line.

c) Finally, in the regime $|\tht_s|<\frac{\hbar\omega}{E}|\tht|$ ($|\qt_s|
<|\qt|$) the soft approximation breaks down, in favor of the high-energy
amplitude~\cite{ACV90}, which also contains emission from the
exchanged-graviton line.
 ~\cite{Li82}.

In the ``soft'' regime a) the amplitude is described by the Weinberg
current~\cite{We65}
[$\eta_i=1(-1)$ for incoming (outgoing) lines]
\begin{eqnarray}\label{JW}
  &&\amp_\soft(E,\Qt; q^\mu)  = \amp_\el J_W^{\mu\nu}\pol_{\mu\nu}
  \equiv  \amp_\el \frac{J_W}{\sqrt{2}} \,, \nonumber \\
  &&J_W^{\mu\nu} = \kappa\sum_i \eta_i \frac{p_i^\mu p_i^\nu}{p_i\cdot q}
  \,,\qquad \kappa \equiv \sqrt{8 \pi G}\,,
\end{eqnarray}
where the complex polarization $\pol^{\mu\nu}$ (together with its complex
conjugate) represents gravitons of definite helicity $\mp2$.  More explicitly we
write $\pol^{\mu\nu}=(\pol_{TT}^{\mu\nu} + \ui\pol_{LT}^{\mu\nu})/\sqrt{2}$,
where  we conveniently fix the gauge by taking:
\begin{eqnarray}
&& \pol_{TT}^{\mu\nu}=\frac{\pol_T^\mu\pol_T^\nu-\pol_L^\mu\pol_L^\nu}{\sqrt{2}}
  \,, \quad
  \pol_{LT}^{\mu\nu}=\frac{\pol_L^\mu\pol_T^\nu+\pol_T^\mu\pol_L^\nu}{\sqrt{2}}
  \,, \\
 && \pol_T^\mu \equiv  \Big(0,-\varepsilon_{ij}\frac{q_j}{|\qt|},0\Big)\,,\;
  \pol_L^\mu \equiv  \frac{(q^3,\bs{0},q^0)}{|\qt|} - (+) \frac{q^\mu}{|\qt|} .
  \nonumber
\end{eqnarray}
for $q$ nearly parallel to $p_1, p_1'$ ($p_2, p_2'$). Concentrating on the former case, 
a simple calculation gives:
\begin{equation}\label{Jc} 
  \frac{J_W }{\sqrt{2}}   = \frac{\kappa E}{\hbar\omega}
  \esp{-2\ui \phi_{\tht}}\left( \esp{2\ui \phi_{\tht-\Tht_i-\tht_s}}
    -\esp{2\ui \phi_{\tht-\Tht_i}} \right)\, .
\end{equation}
Multiplying now eq.~(\ref{Jc}) by the elastic amplitude, we get the following
$b$-space emission amplitude:
\begin{eqnarray}
  &&\tfa_\soft(\bt,E;\omega,\tht) = \sqrt{\frac{Gs}{\hbar}}\frac{R}{\pi}
  \frac{E}{\hbar\omega}  \label{Msoft} \\
  && \qquad \times \int\frac{\dif^2\tht_s}{2\pi\tht_s^2}\;
  \esp{\ui \frac{E}{\hbar} \bt\cdot\tht_s} \, \esp{-2\ui \phi_\tht}
  \frac12\left(\esp{2\ui \phi_{\tht-\tht_s}} - \esp{2\ui \phi_\tht} \right)
  \, , \nonumber
  \end{eqnarray}
  together with its transformation  under change in the incidence angle
$\Tht_i$:
\begin{equation}
\label{transf}
 \tfa_\soft^{(\Tht_i)} = \esp{2\ui(\phi_{\tht-\Tht_i}-\phi_\tht)}
  \tfa_\soft(\bt,E;\omega,\tht-\Tht_i) \, .
\end{equation}

Several remarks are in order. First of all, the current projection shows the
expected $1/\omega$ dependence, but no singularities at either $\tht=\Tht_i$ or
$\tht=\Tht_f = \Tht_i + \tht_s $ as we might have expected from the $p_i\cdot q$
denominators in~(\ref{JW}). This is due to the spin 2 of the graviton, with the
physical projections of the tensor numerators in~(\ref{JW}) providing the result
in terms of scale-invariant phases with azimuthal dependence.  Secondly,
eq.~(\ref{transf}) shows a simple dependence on the incidence angle $\Tht_i$,
which is interpreted as helicity-transformation phase in turning the direction
$\Tht_i$ onto the $\vec{z}$-axis in 3-space, rotation in which the light-like
vector $q^\mu(\tht)$ undergoes the small-angle translation $\tht\to\tht-\Tht_i$.
Finally, the helicity phase transfer in eq.~(\ref{Jc}) can be interpreted by the
$z$-representation ($z\equiv x+\ui y$, $\zt\equiv(x,y)$)
\begin{equation}\label{zRep}
  \esp{2\ui\phi_{\tht_A}}-\esp{2\ui\phi_{\tht_B}}
  = \int\frac{\dif^2 z}{\pi z^{*2}}\,\left(
    \esp{\ui\omega\zt\cdot\tht_B}-\esp{\ui\omega\zt\cdot\tht_A} \right)
\end{equation}
as an integral between initial and final directions in the transverse
$\zt$-plane of the complex component of the Riemann tensor~\cite{GrVe14} in the
AS metric of the incident particles. 

\begin{figure}
  \includegraphics[width=0.99\linewidth]{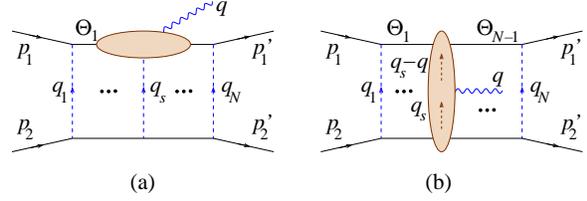}
  \caption{\label{f:channelTransfer} Picture and notation of generic
    exchange emission in (a) the soft and (b) the Regge limits. Here
    $\qt_s=\qt_2$ for the single-exchange amplitude~(\ref{Mregge}).}
\end{figure}

If we now move to larger angles $\theta>\theta_m$, the subenergies increase and
Regge behavior, as described by the Lipatov current~\cite{Li82}
(see also~\cite{ ACV90})
\begin{eqnarray}
  &&
   \frac{J_L^{\mu\nu}}{\qt_{\perp 1}^2 \qt_{\perp 2}^2} \equiv
  \frac{\kappa}{2}\left(\frac{J^\mu J^\nu}{\qt_{\perp 1}^2 \qt_{\perp 2}^2}
    - j^\mu j^\nu\right) ,\,
 j^\mu \equiv \frac{p_1^\mu}{p_1 q} - \frac{p_2^\mu}{p_2 q}
 \nonumber \\
 && J^\mu\equiv \qt_{\perp 1}^2\frac{p_1^\mu}{p_1\cdot q}
 -\qt_{\perp 2}^2\frac{p_2^\mu}{p_2\cdot q} + q_1^\mu-q_2^\mu -\qt_\perp^2 j^\mu
 \,, \label{lip}
\end{eqnarray}
(here the $\qt_\perp$'s are transverse to the $\vp_1$ direction), is turned on.
By performing the same projection as before we get the Regge counterpart of (\ref{Msoft}), (\ref{transf}):
\begin{eqnarray}
  && \tfa_\regge(\bt,E;\omega,\tht) = \sqrt{\frac{Gs}{\hbar}} \frac{R}{\pi}
   \label{Mregge} \\
  && \qquad \times \int\frac{\dif^2\qt_2}{2\pi\qt^2}\;
  \esp{\ui\frac{\qt_2\cdot\bt}{\hbar}} \;
  \frac12\left(1-\esp{-2\ui(\phi_{\qt_2}-\phi_{\qt_2-\qt})} \right)\, ; \nonumber\\
  &&\tfa_\regge^{(\Tht_i)} = \esp{2\ui(\phi_{\tht-\Tht_i}-\phi_\tht)}
  \tfa_\regge(\bt,E;\omega,\tht-\Tht_i)\, , \nonumber
\end{eqnarray}
with the same transformation law as before.  Note that a helicity phase
transfer occurs also in eq.~(\ref{Mregge}), except that the transfer is now in
the $t$-(rather than in the s-)channel (fig.~\ref{f:channelTransfer}).

 Eq.~(\ref{Mregge}) has an explicit representation in terms of $\hti(\bt,\qt)$,
the radiative metric field~\cite{ACV07} and of its $\xt$-space Fourier transform counterpart $ h(\bt,\zt)$:
\begin{eqnarray}
\label{Mreggeh}
 && \tfa_\regge = \sqrt{\frac{Gs}{\hbar}}\frac{R}{2} \, \hti(\bt,\qt) \;; \quad
 \\
 &&    
 h(\bt,\zt) =   \frac{1-\esp{2\ui(\phi_\zt-\phi_{\zt-\bt})}}{2\pi^2 b^2} 
  = \frac{\ui y}{\pi^2 b z^* (z-b)} \nonumber
\end{eqnarray}
as found in part in~\cite{ACV07} and proved in~\cite{CCCV15}. 

Our main observation now is that soft and Regge evaluations agree in the
b)-region. In fact, by setting $\qt_2=\qt_s=E\tht_s$ and $\qt=\hbar\omega\tht$,
we can compute the difference (in complex-$\theta$ notation):
\begin{eqnarray}
  &&\tfa_\regge-\tfa_\soft = \sqrt{\frac{Gs}{\hbar}}\frac{R}{2\pi}
  \frac{E}{\hbar\omega}\int\frac{\dif^2\theta_s}{2\pi|\theta_s|^2}
    \label{DeltaM} \\
  &&\quad\times  \frac{\theta\theta_s^*-\theta^*\theta_s}{|\theta|^2}
    \esp{\ui\frac{E}{\hbar} \bt\cdot\tht_s} \left[
      \frac1{1-\frac{\hbar\omega}{E}\frac{\theta^*}{\theta_s^*}}
      -\frac1{1-\frac{\theta_s^*}{\theta^*}} \right] \,, \nonumber 
\end{eqnarray}
which is clearly vanishing in the angular region
$|\theta|\gg|\theta_s|\gg\frac{\hbar\omega}{E}|\theta|$. This insures the
large-angle regime for the soft amplitude with negligible internal insertions in
the Regge amplitude. Furthermore, eq.~(\ref{DeltaM}) can be used to replace the
soft amplitude with the Regge one in region c) in which the former breaks down.
Since that is a large angle region, the last term can be replaced by $-1$ and
--- with this proviso --- a rescaling of angles in which
$\tilde\theta_s\equiv\frac{E\theta_s}{\hbar\omega}$ is fixed shows that in
region c) eq.~(\ref{DeltaM}) is just the negative of $\tfa_\soft $ at 
$ E = \hbar\omega$.

In conclusion, our matched amplitude, at the single-exchange level, is
\begin{eqnarray}
  \tfa \equiv && 
  \tfa_\soft(\bt,E;\omega,\theta)-\tfa_\soft(\bt,\hbar\omega;\omega,\theta)
  \nonumber \\
  \simeq && -R\sqrt{\frac{Gs}{\hbar}} \frac{\esp{-2\ui\phi_\tht}}{2\pi^2}
  \int\frac{\dif^2 z}{z^{*2}}\;\esp{\ui b\omega\zt\cdot\tht}\Phi(\zt)
  \nonumber \\
  \Phi(\zt)\equiv &&\, \hat{\bt}\cdot\zt+\log|\hat{\bt}-\zt| \, ,\label{Mmatch}
\end{eqnarray}
where the expression on the second line, valid in the $\theta>\theta_m$ region, 
is obtained in a straightforward way~\cite{CCCV15} by using the
$z$-representation~(\ref{zRep}) in the definition~(\ref{Msoft}) of the soft
amplitude. 

We shall call eq.~(\ref{Mmatch}) [eq.~(\ref{Mreggeh})] the soft-based
[Regge-based] representation of the same unified amplitude. Their identity can
be shown~\cite{CCCV15} to be  due to a transversality condition of the
radiative metric tensor~\cite{ACV07} and is thus rooted in the spin-2 nature of the
interaction.

Our next step is to extend the above procedure to any active exchanges in the
eikonal chain, and to resum them, by taking into account two important facts:
{\it(i)} the amplitude transformation [eqs.~(\ref{transf}) and (\ref{Mregge})]
with incidence angle $\Tht_i$ --- possibly much larger that $\theta_m$; and
{\it(ii)} the non-trivial extension of $b$-space factorization to any incidence
angle. The matter is discussed in detail in the parallel paper~\cite{CCCV15},
but the final answer is easy to understand. The resummed single-emission
soft-Regge amplitude, factorized in front of the elastic eikonal $S$-matrix, is
\begin{eqnarray}
  && \ampRid \equiv \frac{\tfa}{\esp{2\ui\delta}}
  = \int_0^1\dif\xi\;\tfa_\regge^{(\xi\tht_s)}
  = \sqrt{\frac{Gs}{\hbar}}\frac{R}{2}  \\
  && \times \int_0^1\dif\xi\;
  \esp{2\ui(\phi_{\tht-\xi\tht_s}-\phi_\tht)}\int\dif^2 z\;
  \esp{\ui b\omega\zt\cdot(\tht-\xi\tht_s)} h(\bt,\zt)   \,. \nonumber
  \label{Mb}
\end{eqnarray}
It represents the coherent average of the single-exchange result of Regge type
at incidence angle $\xi\tht_s$ ranging from 0 to
$\tht_s(\bt)\equiv(2R/b)\hat{\bt}$, with the corresponding transformation phase.
This is even more transparent in an alternative equivalent expression
incorporating the transformation phase
\begin{equation}\label{Mba}
  \frac{-\ampRid}{\esp{-2\ui\phi_\tht}}
  = \sqrt{\frac{Gs}{\hbar}}\frac{R}{\pi}\int_0^1\!\dif\xi\int
  \frac{\dif^2 z}{2\pi z^{*2}}
  \esp{\ui b\omega\zt\cdot(\tht-\xi\tht_s)}\Phi(\zt)
\end{equation}
in which the soft-based form (\ref{Mmatch}) in terms of $\Phi(\zt)$ is used.
The final result~(\ref{Mba}) compares easily with the classical result
of~\cite{GrVe14}: it is the $\xi$-average of the classical amplitude, expanded
to first order in the modulation factor $\Phi(\zt)$.

Our last step consists in using $b$-factorization to resum multi-graviton
emission, at least for the independent pairs, triples, etc.\ of active exchanges
that dominate by combinatorics at high energy. Because of the exponential
counting in eikonal scattering, this provides for us the so-called linear
coherent-state operator in the form
\begin{equation}
  \label{coherent}
  \frac{\hat{S}}{\esp{2\ui\delta}}=\exp \int\frac{\dif^3\vq}{\sqrt{2\omega}}
  2\ui
  \left[\sum_{\lambda=\pm}\ampRid^{(\lambda)}_\bt a_{(\lambda)}^\dagger(\vq\,)
    +\text{h.c.}\right]
\end{equation}
where the helicity amplitude
$\ampRid^{(-)}_\bt(\vq)=[\ampRid^{(+)}_\bt(-\vq,q^3)]^*$ is provided by
eq.~(\ref{Mba}) with a proper identification of variables. Since
operators associated with opposite helicities commute the above coherent-state
is abelian (and thus consistent with the Block-Nordsieck theorem), but describes
both helicities, not only the infrared singular, longitudinal polarization.

\begin{figure}
  \includegraphics[width=\linewidth]{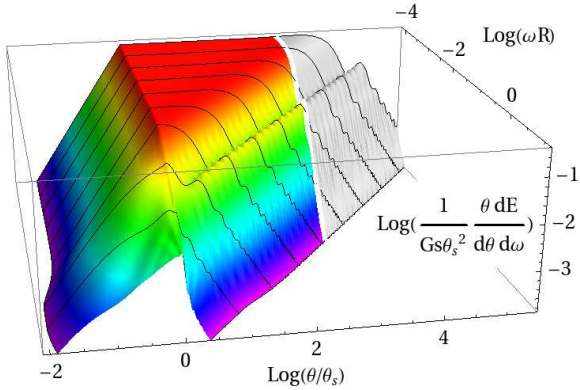}
  \caption{\label{f:resumSpec} Azimuthally integrated spectrum vs.\ $\omega R$
    and $\theta/\theta_s$. The shaded region on the right is excluded by the
    kinematic bound $\theta<1$ (for the choice $\theta_s = 10^{-3}$).}
\end{figure}

\begin{figure}
  \includegraphics[width=\linewidth]{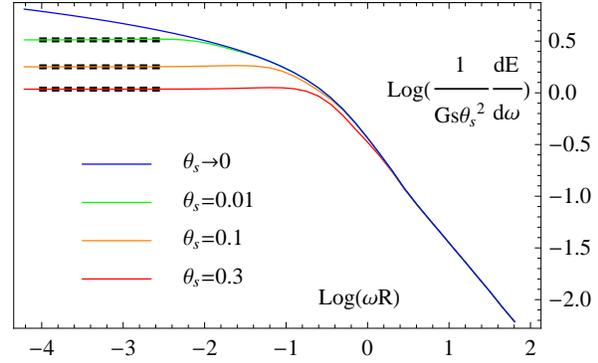}
  \caption{\label{f:freqDist} Frequency spectrum of gravitational
      radiation for various values of $\theta_s$. For each $\theta_s$
      the ZFL value $\frac2{\pi} \log(1.65/\theta_s)$ is obtained (dashed lines).}
\end{figure}

We are finally ready to discuss the gravitational-wave (GW) spectrum. By
normal-ordering in the coherent state (\ref{coherent}) we obtain the
energy-emission distribution
\begin{equation}\label{spectrum}
  \frac{\dif E^{\GW}}{\dif \omega\; \dif \cos \theta\; \dif \phi}
  = \omega^2  \hbar\;2\sum_\lambda|\ampRid^{(\lambda)}_\bt|^2
\end{equation}
The main features of the spectrum can be understood analytically, but can be
best described following the numerical results presented in
fig.~\ref{f:resumSpec} (where we have integrated over the azimuthal angle
$\phi$) and fig.~\ref{f:freqDist} (where we have also integrated over the polar
angle $|\theta|$).

Fig.~\ref{f:resumSpec} shows very clearly that the spectrum is dominated by a
flat plateau (where kinematically accessible) whose shape can be easily
explained as follows. The spectrum falls on the left ($\theta < \theta_s$)
because of phase space and of the absence of collinear singularities. It also
falls when $\omega R = b q \theta_s/\theta > \theta_s/\theta$ since then
$b q >1$. The last limitation (shaded region on the right) is due to the trivial
kinematic bound $\theta<1$.  As a result, for fixed $\omega R < 1$ the length of
plateau in $\log \theta$ is $-\log(\omega R)$ while it disappears completely for
$\omega R >1$. This is the reason why the spectrum in $\omega$ shown in
fig.~\ref{f:freqDist} shows two very distinct regimes:

{\it(i)} $\omega R\ll 1$. In this regime the amplitude~(\ref{Mba}) is well
approximated by dropping the log term in $\Phi(\zt)$ (except if $|b\qt|\gg1$ in
which case, owing to $\Phi(z) \rightarrow z^2$ as $z \rightarrow 0$, the amplitude
behaves as $\sim 1/|b\qt|^2$ (fig.~\ref{f:resumSpec})).  Using the
representation~(\ref{zRep}) and the soft approximation for $\ampRid$, we obtain
the frequency distribution integrated over the whole solid angle
\begin{eqnarray}
 \frac{\dif E^{\GW}}{\dif\omega} =&& \frac{Gs}{\pi}\tht_s^2 \!\int_0^1\!\!
  \frac{2\theta\dif\theta}{\sqrt{1-\theta^2}}\frac{\dif\phi}{\pi}
  \frac{\sin^2\phi_q}{|\tht-\tht_s|^2} \Theta(\frac1{b\omega}-\theta) \nonumber \\
  \simeq && \frac{Gs}{\pi}\tht_s^2
  \left(\!2\log\min\big(\frac{b}{R},\frac1{\omega R}\big)+\text{const} \right)\!.
  \label{freqDist}
\end{eqnarray}
We see that the really infrared regime holds only in the tiny region
$\omega<1/b$, with a rapidity plateau up to $|y|<Y_s\equiv\log(b/R)$
(fig.~\ref{f:resumSpec}), much smaller than $Y=\log(Eb/\hbar)$, the rapidity
available in the single H-diagram emission. On the other hand, here the small-$\omega$ number density in rapidity,
$(Gs/\pi)\tht_s^2$, agrees with the one used in~\cite{ACV90} and with the
zero-frequency limit (ZFL) of~\cite{We65,Sm77}.

{\it(ii)} $1<\omega R< \omega_M R$. In this region the two terms of $\Phi(\zt)$
are of the same order, but for $\omega R\gtrsim 1$ we can look at the integrated
distribution by reliable use of the completeness of the $\qt$-states and we
obtain ($\omega R\gtrsim1$)
\begin{equation}\label{largeOR}
 \frac{\dif E^{\GW}}{\dif\omega}= \frac{2Gs \tht_s^2}{\pi^2}
 \int\frac{\dif^2 z}{|z|^4}\,|\Phi(\zt)|^2 \left(
    \frac{\sin\omega Rx}{\omega Rx}\right)^2
\end{equation}
The spectrum~(\ref{largeOR}) decreases like $1/(\omega R)^2$ for any fixed value
of $x$, in front of an integral which is linearly divergent for $x\to 0$. This
means effectively a $1/(\omega R)$ energy-emission spectrum~\cite{CCCV15}, whose
origin is the decoherence effect induced by $\omega R>1$ on the graviton
emission along the eikonal chain. Furthermore, modulo an overall factor
$\tht_s^2$, the shape of the spectrum is universal above $\omega\simeq b^{-1}$
and tuned on $R^{-1}$ (upper blue curve of fig.~\ref{f:freqDist}).

Thus the total emitted energy fraction is small, of order $\tht_s^2$, and, to
logarithmic accuracy:
\begin{equation}\label{Efrac}
  \frac{E^{\GW}}{\sqrt{s}} \sim \frac23 \log(e/2) \tht_s^2(\bt)
  \log(\omega_M R) \,,
\end{equation}
where $\omega_M$ is an upper frequency cutoff.

Quantum mechanically $\omega_M$ cannot exceed $E/\hbar$ but we expect that the
classical theory ($\hbar \rightarrow 0$) should provide by itself a cutoff.  It
was argued in \cite{GrVe14} that it should correspond to an emission rate
$dE^{\GW}/dt \sim G_N^{-1}$ (sometimes referred to as the Dyson bound) in which
a Planck energy is emitted per Planck time or, classically, the emitted energy
does not have enough time to get out of its own Schwarzschild radius. For our
spectrum falling like $\omega^{-1}$ this gives $R \omega < \tht_s^{-2}$ and
consequently a total fraction of energy loss of order
$\tht_s^{2} \log \tht_s^{-2}$ neatly resolving the ``energy crisis''.

A final question, which deserves attention, is whether we can actually calculate
correlated emission also, by describing the situation in which neighbouring
exchanges act coherently, or a single-exchange emits many gravitons. The latter
correlations originate from multi-H diagrams~\cite{ACV07} and thus contain
higher powers of $R^2/b^2$ in the eikonal, the former ones have been
investigated as rescattering corrections~\cite{CC14}. They may give rise to
higher powers of $\Phi$ and perhaps to the exponentiation proposed
in~\cite{GrVe14} and to a classical large-$\omega$ cutoff, but further
investigation is needed in order to confirm such a guess.

To summarize, we have shown that the spectrum of graviton emission in
transplanckian collisions takes a simple and elegant limiting form, which
unifies the soft and Regge behaviours of the $S$-matrix, and is determined by
the spin-2 structure of the interaction. At low-enough energy the spectrum
reproduces the expected (finite) ZFL. But its shape above $\omega \sim b^{-1}$
is universal and tuned on $R^{-1}$, deviating from the ZFL by a power of
$\omega$ above $\omega \sim R^{-1}$ in agreement with recent classical
results~\cite{GrVe14, Spirin:2015wwa}.  Because of the role of $R^{-1}$, the
characteristic frequency/energy of the emitted gravitons {\em decreases} when
the energy of the collision is {\em increased} above the Planck
scale.

At small deflection angles the radiation is still concentrated around two cones
of size $\ord{\tht_s}$ around the colliding particles, with energies up to
$\ord{R^{-1}}$ and transverse momenta $\ord{b^{-1}}$. Extrapolating
qualitatively the spectrum till $b \sim R$, where classical gravitational
collapse is expected to occur, suggests a smooth quantum transition between the
dispersive and collapsing regimes in transplanckian energy collisions. Such a
smooth transition was already argued to occur in the string-dominated
regime~\cite{Veneziano:2004er} and, more recently, in a $2\to N$ high-energy,
high-multiplicity annihilation process integrated over $\bt$~\cite{Dvali:2014ila}.


\begin{acknowledgments}
  We are grateful to Francesco Coradeschi for several useful discussions. We
  also wish to thank the {\em Galileo Galilei Institute for Theoretical Physics}
  and the {\em Kavli Institute for Theoretical Physics}, University of
  California, Santa Barbara (research supported in part by the National Science
  Foundation under Grant No.\ NSF PHY11-25915) where part of this work was
  carried out.
\end{acknowledgments}

\bibliography{hel}

\end{document}